\documentclass{elsarticle}
\usepackage[letterpaper, margin=1in]{geometry}
\usepackage[colorlinks=true,pdfstartview=FitV,linkcolor=blue,citecolor=blue,urlcolor=blue,bookmarks=false]{hyperref}
\usepackage[usenames,dvipsnames]{color}
\usepackage[sc]{mathpazo}
\usepackage[below]{placeins}
\usepackage{afterpage}
\usepackage[separate-uncertainty,retain-explicit-plus,per-mode = symbol, detect-weight=true, range-phrase=-, range-units=single]{siunitx}
\usepackage{longtable}
\usepackage{multirow}
\usepackage{rotating}
\usepackage[mathlines]{lineno}
\usepackage{graphicx}
\usepackage{paralist}
\usepackage{wrapfig}
\usepackage{appendix}
\usepackage{vruler}
\usepackage{etoolbox}
\usepackage{tikz}
\usepackage{listings}
\usepackage{amsmath}
\usepackage{subfigure}
\usepackage{relsize}
\usepackage{etoolbox}
\usepackage{gensymb}
\modulolinenumbers[5]
\newcolumntype{d}[1]{D{.}{\cdot}{#1}}
\newcommand{\ur}{$^{238}$U}
\renewcommand\th{$^{232}$Th}
\newcommand{\knat}{$^{nat}$K}
\renewcommand{\k}{$^{40}$K}

\newcommand{\ca}{Ca}
\newcommand{\ph}{P}
\newcommand{\dcp}{DCP}
\newcommand{\vov}{v/v}

\newcommand{\dupont}{DuPont}

\DeclareSIUnit\bqkg{Bq/kg_\text{Ar}}
\DeclareSIUnit\ppt{pg/g}
\DeclareSIUnit\ppb{ng/g}
\DeclareSIUnit\ppm{\ensuremath{\micro}g/g}
\DeclareSIUnit\gpg{g/g}
\DeclareSIUnit\c{$c$}
\DeclareSIUnit\day{day}
\DeclareSIUnit\week{w}
\DeclareSIUnit\year{yr}
\DeclareSIUnit\standard{std}
\DeclareSIUnit\str{sr}

\pdfoutput=1
\newif\ifcolorfigs
\colorfigstrue        

\makeatletter
\def\ps@pprintTitle{%
 \let\@oddhead\@empty
 \let\@evenhead\@empty
 \def\@oddfoot{}%
 \let\@evenfoot\@oddfoot}
\makeatother

\journal{}

\biboptions{sort&compress}

\begin{document}
\begin{frontmatter}
\title{Ultra-low radioactivity Kapton and copper-Kapton laminates}

\author{Isaac J. Arnquist\corref{cor1}}
\ead{isaac.arnquist@pnnl.gov}
\author{Chelsie Beck\corref{cor2}}
\author{Maria Laura di Vacri\corref{cor2}}
\author{\\Khadouja Harouaka\corref{cor2}}
\author{Richard Saldanha\corref{cor1}}
\address{Pacific Northwest National Laboratory, Richland, Washington, 99352 USA}
\cortext[cor1]{Corresponding author}
\ead{richard.saldanha@pnnl.gov}
\begin{abstract}
Polyimide-based materials, like Kapton, are widely used in flexible cables and circuitry due to their unique electrical and mechanical characteristics. This study is aimed at investigating the radiopurity of Kapton for use in ultralow background, rare-event physics applications by measuring the \ur, \th, and \knat~ levels using inductively coupled plasma mass spectrometry. Commercial-off-the-shelf Kapton varieties, measured at approximately 950 and 120 \si{\ppt} \ur~ and \th~(1.2$\times$10$^4$ and 490 $\mu$Bq/kg), respectively, can be a significant background source for many current and next-generation ultralow background detectors.  This study has found that the dominant contamination is due to the use of dicalcium phosphate (\dcp), a nonessential slip additive added during manufacturing.  Alternative Kapton materials were obtained that did not contain \dcp~and were determined to be significantly more radiopure than the commercially-available options with 12 and 19 \si{\ppt} \ur~ and \th~(150 and 77 $\mu$Bq/kg), respectively.  The lowest radioactivity version produced (Kapton ELJ, which contains an adhesive) was found to contain single digit \si{\ppt} levels of \ur~and \th, two-to-three orders of magnitude cleaner than commercial-off-the-shelf options.  Moreover, copper-clad polyimide laminates employing Kapton ELJ as the insulator were obtained and shown to be very radiopure at 8.6 and 22 \si{\ppt} \ur~ and \th~(110 and 89 $\mu$Bq/kg), respectively.
\end{abstract}

\end{frontmatter}
\section{Introduction}
Signal sensors and their associated cabling and readout electronics are often a significant contributor to the radioactive background budget of rare-event experiments such as searches for neutrinoless double beta decay or the direct detection of dark matter. \cite{kharusi2018nexo, cebrian2017radiopurity, busch2018low, andreotti2009low, agnese2017projected, aguilar2016search, armengaud2017performance}. Circuitry and cables are typically composed of two major components - the conductor (typically copper) and insulator. Kapton is a polyimide that is widely used as an insulating substrate in the electronics industry due to its unique properties of high resistivity, high dielectric strength, and flexibility. It is also stable across a wide range of temperatures, has good thermal conductivity, a thermal expansion coefficient that is close to copper, and a low outgassing rate, which make it a favorable material for use in the ultra-high vacuum and cryogenic environments that are commonly found in low background experiments. 

While extremely radiopure copper can be obtained \cite{abgrall2016majorana}, commercial Kapton is not a very radiopure material, with measured contamination levels of roughly 1400 ppt  \ur~ (1.7$\times$10$^4$ $\mu$Bq/kg) \cite{nisi2009comparison}, leading to high-radioactivity components. Low-background experiments therefore have to either limit the amount of Kapton used to the absolute minimum necessary \cite{kharusi2018nexo}, or use other materials \cite{budjavs2009gamma, andreotti2009low, busch2018low} that, while more radiopure, do not have all the advantageous properties of Kapton. Additionally, since Kapton is an industry standard for electronics, the use of alternative material often requires custom-made components, which can increase costs, risk, and production time. 
Sourcing a radiopure Kapton at levels 100-1000x cleaner than the commercially-available options would allow for significant reduction in the radioactive backgrounds and would also reduce the constraints on signal sensors and readout, potentially increasing the overall sensitivity of several low-background experiments.  

There is nothing intrinsically radioactive about the chemical composition of Kapton; it is an organic polymer composed of H, C, O, and N. It is therefore plausible that radiopure Kapton could be sourced at ultrapure levels (\textit{e.g.}, parts-per-trillion or \si{\micro\becquerel\per\kg} levels), as seen for polymethylmethacrylate ("acrylic"), polyvinylidene fluoride (PVDF), bioabsorbables, polyetherimides (PEI, trade name ULTEM), polychlorotrifluoroethylene (PCTFE), \textit{etc}. \cite{arnquist2017mass, arnquist2019mass}. However, polymer process engineering oftentimes involves the use of additives or processing reagents that could be vectors of contamination (\textit{e.g.}, fillers, inorganic synthesis reagents, catalysts, crosslinkers, and additives) and/or contaminants could be imparted through handling, rolling, machining, and other physical manufacturing procedures.  For example, in PVDF process engineering, radiopure stock powder has been shown to be quite pure ($\approx$ \si{\micro\becquerel\per\kg}), but gets more contaminated after each processing step (\textit{e.g.,} from powder to pellet to final formed part) \cite{arnquist2017mass}. For Kapton, if the contaminating step(s) could be identified, controlled, and/or mitigated, perhaps films could be obtained at significantly reduced activity levels.

The intent of this study was to source radiopure Kapton for low radioactivity flexible cables and circuits. Investigations were made into understanding backgrounds from a variety of Kapton variants through ultra-sensitive assay.  Production and process engineering steps were researched to understand where vectors of contamination could be introduced, and then accounted for and mitigated. Once radiopure Kapton was identified, copper-Kapton laminates were tested to determine whether radiopurity was maintained during the lamination process.

\section{Analysis Methods}
\label{sec:analysis}
All analyses of the materials discussed in this paper were performed at Pacific Northwest National Laboratory (PNNL). A laminar flow hood providing a Class 10 environment was used for sample preparations while all other experimental work was performed in a Class 10000 cleanroom. Details on the chemical reagents used and the preparation of all labware prior to sample handling is given in \ref{sec:labware}.

\subsection{Kapton cleaning, digestion, and dry ashing}
\begin{figure}
    \centering
    \includegraphics[width = \linewidth]{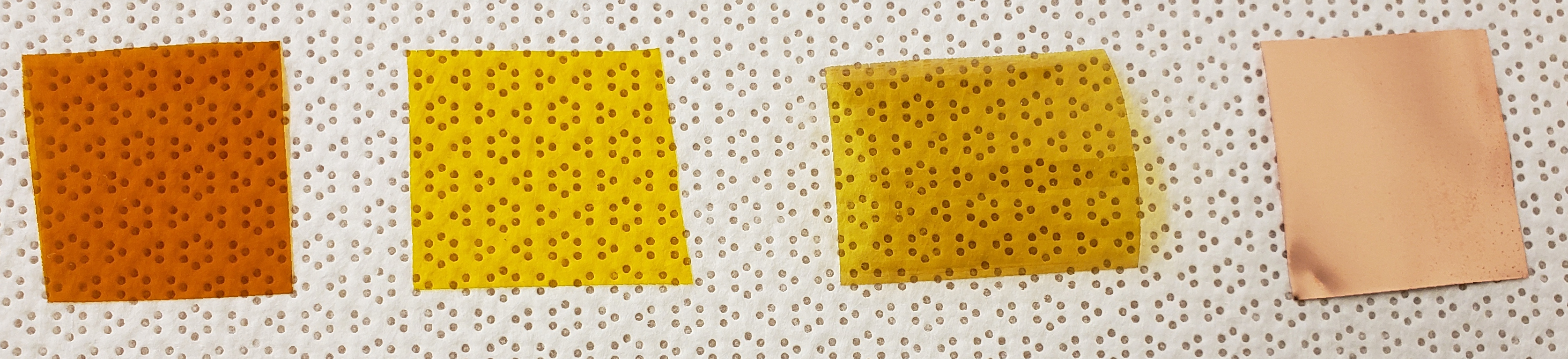}
    \caption{Samples of the Kapton films and Kapton-copper laminates analysed for this study. From left to right: Kapton 300 HN, Kapton 200 HH, Kapton 300 ELJ, Kapton 300ELJ + Copper laminate}
    \label{fig:all_samples}
\end{figure}

Subsamples on the order of 50 mg were cut from each Kapton foil using clean stainless steel scissors. In order to remove surface grease and contamination residues from handling and cutting operations, a cleaning procedure was applied prior to digestion. Subsamples were sonicated in 2\% \vov~Micro-90\textsuperscript{\textregistered} detergent solution at $\SI{45}{\celsius}$ for 25 min, followed by rinsing in MilliQ water and sonication in 6M Optima grade HNO$_3$ solution at \SI{45}{\celsius} for 20 min. After rinsing in MilliQ water, samples were then sonicated in MilliQ water at \SI{45}{\celsius} for 15 min, followed by a final rinse with MilliQ water and air dried in a Class 10 laminar flow hood.

Subsamples were transferred, using low background clean plastic tongs, to validated PTFE Mars 6 iPrep$^{TM}$ microwave digestion vessels. All samples and process blanks were spiked with a known amount, on the order of \SIrange{100}{200}{\femto\gram}, of $^{229}$Th and $^{233}$U radiotracers (Oak Ridge National Laboratory, Oak Ridge, TN) before digestion. The amount of spiked tracer was gravimetrically measured. Triplicates of each sample and three process blanks were prepared for each batch of analysis. 

A Mars 6 microwave digestion system (CEM corporation, Charlotte, NC), equipped with 12 iPrep$^{TM}$ polytetrafluoroethylene (PTFE) digestion vessels, was used for microwave assisted digestion of the samples at $\SI{250}{\celsius}$ in 5 mL of concentrated Optima grade HNO$_3$. After complete digestion, sample solutions were transferred to cleaned and validated PFA Savillex vials (Bloomington, MN). Concentrated acid was boiled off on a hotplate at \SI{170}{\celsius}, samples were reconstituted in 2$\%$ Optima grade nitric acid solution.

For ultrasensitive K analyses, we have found that microwave digestion vessels and standard quartz crucibles do not provide the consistently low K backgrounds required to reach the sensitivity needed. A dry ashing technique employing ultraclean electroformed copper (EFCu) crucibles, providing sufficiently low K backgrounds, was adapted from \cite{arnquist2017mass}. In brief, the subsamples were first weighed in ultralow background EFCu crucibles and were then dry ashed overnight in a programmable tube furnace in a flowing air (\SI{4} {\liter\per\min}) atmosphere, reaching a maximum temperature of \SI{800}{\celsius}.  The sample and process blank crucibles (along with any residue from the ashing process) were retrieved, digested to 1000 ppm Cu and assayed for K using one of our triple quadrupole ICP-MS instruments.     
\subsection{ICP-MS Analysis} 
Determinations of Th, U, and K were performed using either an Agilent 8800 or 8900 ICP-MS (Agilent Technologies, Santa Clara, CA), each equipped with an integrated autosampler, a microflow PFA nebulizer and a quartz double pass spray chamber. Plasma, ion optics and mass analyzer parameters were adjusted based on the instrumental response of a standard solution from Agilent Technologies and/or an in-house K standard. For U and Th analysis, the instrument was tuned to maximize sensitivity in the high $m/z$ range, at the expense of lower $m/z$ signals, in order to optimize the signal-to-noise ratio for Th and U. The instrumental response for Tl from the \SI{0.1}{\ppb} standard tuning solution was used as a reference signal. Oxides were monitored and kept below 2$\%$ based on the $m/z = $ 156 and $m/z = $ 140 ratio from Ce (CeO$^{+}$/Ce$^{+}$) in the tuning standard solution. An acquisition method of three replicates and ten sweeps per replicate was used for each reading. Acquisition times for monitoring $m/z$ of interest (\textit{e.g.,} tracers and analytes) were set based on expected signals, in order to maximize instrumental precision by improving counting statistics. \\
Quantitation of $^{232}$Th and $^{238}$U was performed using isotope dilution methods, using the equation:

\begin{equation}
    \text{Concentration} = \dfrac{A_\text{analyte} \cdot C_\text{tracer}}{A_\text{tracer}} 
\end{equation}
where $A_\text{analyte}$ is the instrument response for the analyte, $A_\text{tracer}$ is the instrumental response for the tracer and $C_{tracer}$ is the concentration of the tracer in the sample.  Quantitation of K was performed using an external calibration curve in matrix matched standards.  Absolute detection limits on the order of 20 \textit{femto}grams (\si{fg}) is common using the above method for \ur~and \th, while absolute detection limits of 1 \textit{pico}gram (\si{pg}) is common for \knat.  These sensitivities offer sub-microBq/kg determinations on samples on the order of only 100 mg.
All central values and uncertainties reported are the average value and standard deviation of three independent replicates respectively. For laminate samples (Section \ref{sec:laminates}), six independent replicates were measured and reported values are the average and standard deviation of the six replicates. Samples for which the analyte concentration was below the detection limit are reported as an upper limit. Values are reported in pg/g of sample for $^{238}$U and $^{232}$Th, and in ng/g of sample for $^{nat}$K. They can be converted to $\mu$Bq/kg of sample based on the specific activity and isotopic abundance of the radionuclides. For reference, 1 pg $^{238}$U/g of sample corresponds to 12.4 $\mu$Bq/kg of sample, 1 pg $^{232}$Th/g of sample corresponds to 4.06 $\mu$Bq/kg of sample, and 1 ng $^{nat}$K/g of sample corresponds to 30.5 $\mu$Bq/kg of sample from $^{40}$K.  

\subsection{ICP-OES Analysis} 
Determinations of Ca and P were performed using a Thermo iCAP7600 ICP Optical Emission Spectrometer (OES) Duo (Thermo Scientific, Waltham, MA) equipped with standard quartz sample introduction. The instrument was calibrated using dilutions of 1000 ppm single element standards (Inorganic Ventures, Christianburg, VA) in 2\% HNO$_3$. Calcium was calibrated from \SIrange{5}{500}{\ppb} and phosphorus from \SIrange{0.1}{10}{\ppm}. The regression coefficient for both calibration curves was $>$0.9999. Multiple wavelengths for both Ca and P were chosen and compared to verify the absence of spectral interferences. The highest intensity (most sensitive) wavelengths were used for quantitation.  
\section{Commercial Kapton}

As a baseline for our work, commercially-available off-the-shelf Kapton HN films were assayed for the key radioimpurities relevant to low background experiments (\ur, \th, \k). Table~\ref{tab:commercial_kapton} shows the determinations for four films sourced from \dupont~and one film sourced from a distributor \cite{cshyde}, ranging in thickness from 1-5 mil. The measured contamination of \ur~is fairly consistent across all samples with an average value of $\approx$ \SI{950}{\ppt}. There is slightly more variation in the measured \th~contamination across samples with an average of $\approx$ \SI{120}{\ppt}. Due to time constraints only one sample was measured for potassium, with the \knat~concentration $\approx$ \SI{40}{\ppb}. These values are in fairly good agreement with previous measurements in the literature (see last row of Table~\ref{tab:commercial_kapton} \cite{nisi2009comparison}).

The overall activity levels for \ur~and \th~are relatively high for current and next-generation rare event searches, and would be a dominant and sensitivity-limiting background source for many proposed experiments.  

\begin{table}[t]
\centering
\begin{tabular}{c c c c c c}
\hline
\textbf{Type} & \textbf{Thickness [\si{mil}]} & \textbf{Vendor}  & \textbf{\ur~[\si{\ppt}]} & \textbf{\th~[\si{\ppt}]} & \textbf{\knat~[\si{\ppb}]}\\ 
\hline
100HN &  1 & \dupont & \num{966 \pm 14} & \num{89.8 \pm 3.2} & \\ 
100HN & 1 & \dupont & \num{931.1 \pm 5.2} & \num{115.0 \pm 1.2} &\\ 
300HN & 3 & \dupont & \num{1082 \pm 43} & \num{249.6 \pm 8.5} & \num{44 \pm 18}\\ 
500HN & 5 & \dupont  & \num{826 \pm 73} & \num{72.9 \pm 6.5} &\\ 
500HN & 5 & CS Hyde & \num{921 \pm 11} & \num{86.5 \pm 1.3} &\\ 
\hline
$^*$HN & & \dupont & \num{1400 \pm 160} & \num{160 \pm 20} & \num{200 \pm 100}\\
\hline
\end{tabular}
\caption{Radioassay measurements of commercial off-the-shelf Kapton HN films with different thicknesses. $^*$The last row shows previously published measurements \cite{nisi2009comparison} for comparison.}  
\label{tab:commercial_kapton}
\end{table}

\section{Understanding Kapton Process Engineering} 
It is worth noting that it is extremely uncommon that materials are directly contaminated with sources of \textit{pure} uranium and/or \textit{pure} thorium.  Instead, \ur~and \th~tag along as concomitant impurities in the host material, be it through added reagents, mechanical exposure, dust particulates, \textit{etc.}, as discussed in Section 1. In order to understand how radioactive contamination enters a material it is important to understand how it is manufactured. Through discussions with technical representatives at \dupont~and Fralock (makers of Cirlex, a laminated form of Kapton) we learned of a slip additive that is added to commercially-available Kapton. 

In the early 1980s \dupont~introduced a new version of their standard Kapton H product, referred to as Kapton HN, in order to improve film handling \cite{williams1990kapton, mcalees2002raw}. The new Kapton HN had the same chemistry as the original Kapton H, but contained a slip additive: calcium phosphate dibasic (CaHPO$_4$, or, colloquially, dicalcium phosphate) \cite{williams1990kapton, wang1988characterization}, hereafter referred to as \dcp. The addition of a slip additive into the material is intended to reduce friction between film layers during the production process. Slip additives are typically added homogeneously to the polymer in the molten phase. As the melt begins to solidify, the slip additive tends to agglomerate (up to a few microns in size \cite{williams1990kapton, wang1988characterization, hin2009materials}) and the majority of the slip additive particles migrate from the bulk matrix to the surface of the film \cite{keck2010additives}. The presence of the slip additive particles on the film surface reduces direct film-to-film contact which can prevent film sticking and pulling, helping to increase film production throughput \cite{hin2009materials}.

The chemical composition and concentration of the slip additive DCP ($\approx$ 1000 ppm or 0.1\%) made it a prime candidate for being the dominant contamination pathway for \ur~and \th. Phosphates (along with carbonates and sulfates) are notorious for forming sparingly soluble metal complexes with most metal cations (\textit{sans} alkali earth metals, like Li$^+$, Na$^+$, K$^+$, \textit{etc.}).  Moreover, phosphate minerals are known for having high amounts of naturally-occurring radioactive materials \cite{ncrp1986radiation}.   

\begin{table}[t]
\centering
\begin{tabular}{c c c c c c}
\hline
\textbf{Type} & \textbf{Thickness [mil]} & \textbf{Source} & \textbf{\ur~[\si{\ppt}]} & \textbf{\th~[\si{\ppt}]} & \textbf{\knat~[\si{\ppb}]} \\ 
\hline
200HH & 2 & R\&D line & \num{12.3 \pm 1.9} & \num{18.5 \pm 2.3} & \num{34 \pm 14}\\ 
100ELJ & 1 & R\&D line & \num{2.42 \pm 0.76} & $<$ 4.5 & \num{71 \pm 36} \\ 
300ELJ & 3 & R\&D line & \num{5.58 \pm 0.16} & \num{6.82 \pm 0.09} & \num{108 \pm 38} \\
\hline
\end{tabular}
\caption{ICP-MS radioassay results for Kapton samples produced without \dcp~(received from the \dupont~R\&D division). The HH films are alternatives to the standard HN films while the ELJ films are prepared for copper lamination.}
\label{tab:dcp_less}
\end{table}

As a test of this hypothesis, we requested Kapton film samples from our \dupont~contacts without the \dcp~additive, as was the original formulation for early Kapton H films. These samples, referred to as Kapton HH, are no longer commercially available and were specially produced for us by \dupont~at an R\&D facility. The 2-mil thick 200HH sample films were assayed in the same way as the commercial Kapton HN films and were found to have contamination levels of \num{12} and \SI{19}{\ppt} for \ur~and \th, respectively, roughly factors of 77x and 6x lower than that of commercial Kapton HN (see Table~\ref{tab:dcp_less}). The level of \knat~was consistent with the measured values for Kapton HN.

\begin{table}[t]
\centering
\begin{tabular}{c c c c}
\hline
\textbf{Type} & \textbf{Source} & \textbf{\ca~[\si{\ppm}]} & \textbf{\ph~[\si{\ppm}]} \\ 
\hline
100HN & Production & 352.4 $\pm$ 1.3 & 289.9 $\pm$ 1.3 \\ 
500HN & Production & 318 $\pm$ 14 & 259 $\pm$ 11 \\ 
200HH & R\&D & $<$ 0.28 & $<$ 0.27 \\ 
100 ELJ & R\&D & $<$ 0.68 & 23.34 $\pm$ 0.32 \\
\hline
\end{tabular}
\caption{ICP-OES radioassay results for \ca~and \ph~concentrations in Kapton samples from \dupont.}
\label{tab:icpoes}
\end{table}

To confirm that the Kapton HH did in fact have less \dcp~than the commercial Kapton HN, samples of both film types were analyzed for \ca~and \ph~using ICP-OES. The results, shown in Table~\ref{tab:icpoes}, indicate the presence of roughly \SI{335} and \SI{275}{\ppm} for \ca~and \ph~ in Kapton HN, respectively.  The ratio of \ca~and \ph~is consistent with the ratio expected for DCP (\ca:\ph~$\approx 1.3$ by weight) and would correspond to a concentration of roughly 0.12\% DCP in the Kapton HN, which is again consistent with the expected amount of slip additive in the commercially-available HN films. The HH film, on the other hand, had \textit{no} detectable amounts of either \ca~or \ph, at least a factor of 1000x lower in concentration than that seen in the HN films.

Finally, we also obtained a sample of the \dcp~slip additive in the form of a fine powder. The \dcp~sample was analyzed via ICP-MS resulting in measured values of  \SI{1.5}{\ppm} and \SI{0.079}{\ppm} for \ur~and \th,  respectively. These measurements confirm the notion that phosphates can have relatively high concentrations of \ur~and \th, and also accounts for the determinations of \ur~and \th~ seen in commercial Kapton HN films with 0.1\% \dcp~. 

To summarize, through measurements of the \ur~and \th~contamination in \dcp~and the \dcp~ concentration in Kapton HN films we have conclusively identified the dominant contamination vector of \ur~and \th~in commercial Kapton HN films to be the slip additive CaHPO$_4$ (\dcp).  Additionally, we have shown that Kapton films made \textit{without} \dcp~have significantly lower \ur~and \th~contamination levels. It should be noted that the addition of \dcp~is primarily for increasing film production throughput and Kapton HH, produced without \dcp, has the same advantageous electrical and mechanical properties as commercially available Kapton HN \cite{williams1990kapton}.





\section{Laminates}
\label{sec:laminates}
While we have shown that radiopure Kapton film can be produced, there are several further manufacturing steps involved before these films can be used in flexible cables or circuitry. In a standard production process the Kapton is first layered with a conductive metal such as copper, typically with the use of an adhesive, to form a laminate. The desired electrical layout is then patterned on the laminate in the form of a resist using screen printing or photo-lithography. Unexposed copper regions are then etched away before the resist is removed. Finally through holes and vias are made to connect different conductive layers. Each one of the above steps involves a large amount of chemical processing, all of which could potentially add radioactive contaminants to the final product. It is therefore extremely important that each step of the manufacturing process is carefully controlled. In this paper we have studied the first steps of this manufacturing process - the addition of polyimide adhesive to Kapton film and the subsequent lamination with copper.

\subsection{Adhesive Layer} 
In order to ensure strong and stable bonding of the Kapton substrate to the copper layer, an adhesive layer is often used. The adhesive is generally coated onto the Kapton substrate before it is laminated to copper foil using pressure and heat. In order to investigate if the addition of the adhesive layer introduces radioactive contaminants, we obtained samples of Kapton ELJ from \dupont. Kapton ELJ is a coated polyimide film that consists of a Kapton E core coated on each side with a layer of Kapton LJ. Kapton E is a film specifically developed for flexible circuitry applications. It consists of a mix of two dianhydrides, PMDA (pyromellitic dianhydride) and BPDA (biphenyltetracarboxylic acid dianhydride), and two diamines, ODA (oxydiphenylene diamine) and PPD (paraphenylenediamine). Kapton E has a lower coefficient of thermal expansion (better matched to copper), reduced moisture absorption and reduced coefficient of hydroscopic expansion \cite{mcclure2010polyimide}. Kapton LJ is a thermoplastic copolyimide film derived from 80 to 95 mole\% 1,3 bis(4-aminophenoxy) benzene, 5 to 20 mole\% hexamethylene diamine and 100 mole\% 4,4 oxydiphthalic dianhydride \cite{lee2000high} which is used as a low temperature polyimide adhesive.

We obtained two samples of Kapton ELJ produced without the addition of DCP from the Dupont R\&D line: 100ELJ and 300ELJ with 1 and 3 mil thicknesses, respectively. The samples were analyzed for \ur, \th, and \knat~using ICP-MS as well as for \ca~and \ph~using ICP-OES, with the results shown alongside previously mentioned variants of Kapton in Tables~\ref{tab:dcp_less} and~\ref{tab:icpoes}, respectively. It can be seen that the \ur~and \th~contamination levels are even further reduced compared to the 200HH film, down to the single-digit ppt levels. Kapton ELJ has reduction factors of roughly 200x and 20x for \ur~ and \th, respectively, compared to typical commercial-off-the-shelf Kapton HN variants. The level of \knat~is slightly increased to roughly \SI{90}{\ppm}. Interestingly, while the 100 ELJ film showed no detectable levels of \ca, roughly \SI{20}{\ppm} of \ph~was detected, indicating the presence of a different phosphorous-based chemical that, unlike DCP, does not introduce significant \ur~or \th~contamination. 

\subsection{Kapton-Copper Laminate}
Following the encouraging results above which show that the addition of the adhesive layer does not introduce contaminants, we proceeded to investigate laminates of Kapton and copper. We obtained a laminate sample from \dupont~that consisted of 3 mil thick DCP-less 300ELJ Kapton foil (the same as the one described in Section 5.1), sandwiched between roll annealed copper foil (\SI{0.5}{oz\per ft\squared} on each side). Analysis results are shown in Table \ref{tab:laminate}. Contamination levels of \ur,~\th,~and \knat~show significantly larger variations (the uncertainties are derived from variations in the results obtained from different subsamples) and higher central values in the finished laminate compared to the starting 300ELJ Kapton.

The next step was to investigate the source of the increased variability in contamination of the laminate and explore the possibility of obtaining laminates as pure as the starting Kapton foil. Since the major component added to Kapton to manufacture the laminate is copper, this material was investigated as a possible carrier of contamination. The copper fraction of the 300ELJ laminate was selectively dissolved and analyzed by ICP-MS for \ur, \th, and \knat. After a mild cleaning of copper-polyimide subsamples in 2$\%$ Micro90 detergent followed by rinsing in MilliQ water and air drying in a Class 10 laminar flow hood, complete dissolution of the copper layers was performed by etching the copper away from the subsample laminates in Optima grade 8M HNO$_3$ . The obtained copper solutions were then diluted and analyzed via ICP-MS. 

Results for \ur, \th, and \knat~in the copper component of the laminate are reported in Table~\ref{tab:laminate}. The values are normalized to the \textit{total mass of sample} (not just copper) to allow for a direct comparison of contamination from copper relative to determinations for the whole polyimide-copper laminate.  When normalizing impurities from the copper relative to the total mass of laminate, it is apparent that the variability in total radioactivity of the laminate may be due to the copper layers.

Table~\ref{tab:copper} shows a direct comparison of determinations in Kapton ELJ and copper from Kapton ELJ laminates, with values shown normalized to the Kapton and copper, respectively. These data directly show the high variability across copper sample replicates (relatively high standard deviation compared to Kapton ELJ). This clearly indicates that the copper layers do not have a uniform distribution of contaminants across the material, either due to the intrinsic variability of the source copper and/or from the copper processing/lamination step(s).  While it is not uncommon to find commercially-available copper with sub-ppt levels of \ur~ and \th~ for ultralow background applications, such levels are more typical of "bulk" copper with small surface area-to-volume ratios.  It is much less common to find thin layers of copper with radiopurities at those levels due to the higher surface area and handling.  With that said, assuming the ideal scenario of the copper being a negligible contributor to the radioactivity of the laminate (\textit{i.e.,} all the radioactivity is coming from the Kapton ELJ), the overall contamination levels of \ur~ and \th~ in the final ideal laminate would be on the order of \SIrange{1}{2}{\ppt}, about an order of magnitude better than the already impressive, values shown in Table~\ref{tab:laminate}.   

\begin{table}[ht]
\centering
\begin{tabular}{c c c c c c}
\hline
\textbf{Type} & \textbf{Polyimide} & \textbf{Copper } & \textbf{\ur} & \textbf{\th} & \textbf{\knat}\\ 
 & \textbf{Thickness} & \textbf{Thickness} &  & &  \\
  & \textbf{[\si{\micro\meter}]} & \textbf{[\si{\micro\meter}]} & \textbf{[pg/g laminate]} & \textbf{[pg/g laminate]} & \textbf{[ng/g laminate]} \\
\hline
300ELJ+Cu laminate & 76.2 & 17.0 (x2) & \num{8.6 \pm 3.6} & \num{20 \pm 14} & \num{164 \pm 82} \\
Copper only & - & - &\num{8.6 \pm 9.6} & \num{8.7 \pm 8.6} & \num{100 \pm 130}\\
\hline
\end{tabular}
\caption{$^{238}$U, $^{232}$Th and $^{nat}$K contamination measured in the 300ELJ roll annealed copper laminate. The contribution measured in the copper fraction of the laminate, normalized to the original mass of \textit{copper-polyimide laminate}, is also reported.}
\label{tab:laminate}
\end{table}
\begin{table}[t]
\centering
\begin{tabular}{c c c c c}
\hline
\textbf{Material} & \textbf{Mass Fraction} & \textbf{\ur} & \textbf{\th} & \textbf{\knat} \\ 
 & \textbf{in Laminate [\%]} & \textbf{[\si{\ppt}]} & \textbf{[\si{\ppt}]} & \textbf{[\si{\ppb}]} \\ 
\hline
Kapton 300ELJ & 26  & \num{5.58 \pm 0.16} & \num{6.82 \pm 0.09} & \num{108 \pm 38} \\
Copper from 300 ELJ Laminate & 74 & \num{12 \pm 13} & \num{12 \pm 12} & \num{130 \pm 180} \\
\hline
\end{tabular}
\caption{ICP-MS radioassay results from Kapton ELJ and the copper removed from Kapton ELJ laminates normalized to Kapton and copper, respectively}
\label{tab:copper}
\end{table}

In order to compare the levels of radioactive contamination measured in the DCP-less laminates to commercially available laminates, three samples of laminates from different vendors were procured and analyzed. The laminates were digested following the same digestion procedure described in Section~\ref{sec:analysis} and analyzed for \ur~and \th~using ICP-MS. The results are reported in Table~\ref{tab:commercial_laminates}.  As can be seen, the \ur~contamination in the Kapton ELJ laminate is about a factor of 20x cleaner than a commercially-available Pyralux variety with the same specifications, and much cleaner than Novaclad and Cirlex alternatives. The \th~and \knat~contamination levels in the Kapton ELJ laminate are compatible with the commercially available Pyralux laminate.

\begin{table}[t]
\centering
\begin{tabular}{c c c c c c c c}
\hline
\textbf{Type} & \textbf{Vendor} & \textbf{Polyimide} & \textbf{Copper } & \textbf{Polyimide} & \textbf{\ur} & \textbf{\th} & \textbf{\knat} \\ 
 & & \textbf{Thickness} & \textbf{Thickness} & \textbf{Frac.} & & & \\
  & & \textbf{[\si{\micro\meter}]} & \textbf{[\si{\micro\meter}]} & \textbf{[\%]} & \textbf{[\si{\ppt}]} & \textbf{[\si{\ppt}]} & \textbf{[\si{\ppb}]} \\
\hline
300ELJ+Cu laminate & \dupont~R\&D & 76.2 & 17.0 (x2) & 26 & \num{8.6 \pm 3.6} & \num{20 \pm 14} & \num{164 \pm 82} \\
\hline
Pyralux AP8535R & \dupont & 76.2 & 17.0 (x2)& 26 & \num{158.0 \pm 6.1} & \num{24.1 \pm 0.9} & $<$ 210\\ 
Novaclad 146319-009 & Sheldahl & 50.8 & 5.0 (x1) & 62 & \num{283 \pm 21} & \num{50.1 \pm 3.9} & $<$ 210\\
Cirlex & Fralock & 228.6 & 34.1 (x2) & 35 & \num{413 \pm 45} & \num{71.4 \pm 2.1} & $<$ 210\\
\hline 
\end{tabular}
\caption{Comparison of the special DCP-less copper-polyimide laminate (top row) with commercially available options. The (xN) factor  in the fourth row indicates whether the copper was on one or both sides of the laminate.}
\label{tab:commercial_laminates}
\end{table}

\section{Summary}
Investigations into the search for radiopure Kapton provided some very positive results.  Through our study, the contamination vector of commercially-available Kapton HN was identified to be stemming from the use of DCP, an optional slip additive customarily employed at $\sim$ 1000 ppm levels in commercially-available Kapton. Versions of Kapton were identified that are significantly cleaner than commercially-available Kapton HN varieties, providing \ur~and \th~contamination levels in the low parts-per-trillion regime.  Reductions factors as high as 393x and 27x were determined for \ur~and \th~compared to Kapton HN varieties. Investigations into clean Kapton-copper laminates also showed promising results, with a Kapton ELJ laminate found at 8.6 and 20 \si{\ppt} for \ur~and \th~(110 and 81 $\mu$Bq/kg), respectively.  Interestingly, selective separate assays of the polyimide (Kapton) and roll annealed copper layers revealed a higher and non-homogeneous distribution of contaminants in the latter, as shown by the significantly high uncertainties among replicates relative to the central value (Table~\ref{tab:copper}).  

Further investigations (currently underway) will focus on three areas of study.  First, laminates employing electrodeposited copper layers (compared to roll annealed copper) will be investigated to determine if laminates can be made even cleaner through a alternative application of copper. Secondly, similar studies will be applied to investigating and sourcing radiopure Cirlex, a thicker laminate form of Kapton, which is a very useful, albeit relatively radioactive, flexible circuit material employed in many rare event detectors. Lastly, the final step of cable making, the lithography, will be investigated to identify and mitigate background sources introduced through the processing of laminates into cables.  A "start clean, stay clean" approach will be followed in order to go from clean Kapton to laminate to finished cable.
\section{Acknowledgments}
The authors are extremely grateful to Mark McAlees and Rosa Gonzalez at \dupont~for informative discussions on the Kapton production process and for providing numerous samples that allowed for the detailed analyses presented in this paper. We would also like to acknowledge Marcelo Norona at Fralock for helpful discussions and David Moore at Yale University for providing the commercial laminates.  This work was funded by PNNL Laboratory Directed Research and Development funds under the  Nuclear Physics, Particle Physics, Astrophysics, and Cosmology Initiative.  The Pacific Northwest National Laboratory is a multi-program national laboratory operated for the U.S. Department of Energy (DOE) by Battelle Memorial Institute under contract number DE-AC05-76RL01830.  
\bibliography{references.bib}
\appendix
\section{Reagents and Labware}
\label{sec:labware}
Micro-90\textsuperscript{\textregistered} detergent (Cole-Parmer, Vernon Hills, IL) was employed for sample cleaning. Optima grade nitric acid was used for sample cleaning and preparation, 18.2 M$\Omega\cdot$cm water from a MilliQ system (Merk Millipore GmbH, Burlington, MA) was used for sample rinsing and in the preparation of reagent solutions. Ultralow background perfluoroalkoxy alkane (PFA) screw cap vials from Savillex (Eden Prairie, MN) were used as sample containers, to collect solutions from microwave digestion vessels and as ICP-MS autosampler vials. \\
All labware involved in the sample handling and analysis (vials, microwave vessels, tongs, pipette tips) were cleaned with 2\% \vov~Micro-90 detergent, triply rinsed with MilliQ water and leached in Optima grade 3M HCl and 6M HNO$_3$ solutions. Following leaching, all labware underwent a validation to ensure cleanliness. The validation step consisted of pipetting a small volume of 5\% \vov~HNO$_3$ into each container, 1.5 mL in the PFA vials, 5mL in the iPrep$^{TM}$ microwave vessels. Vials were closed, shaken, and kept at $\SI{80}{\celsius}$ for at least 12 hours. Microwave vessels underwent a microwave digestion run at $\SI{220}{\celsius}$. Tongs and pipette tips were soaked into a 5\% \vov~HNO$_3$ leaching solution ($ca.$ 1.5 mL) for few minutes. The leachate from all labware was then analyzed via ICP-MS. The validation was performed to assure sufficiently low background for Th and U. Only labware for which Th and U signals were at reagents background levels passed validation. Microwave vessels were validated also for Ca. Labware failing validation underwent additional cycles of leaching and validation tests until meeting background requirements. 
\end{document}